\documentclass[journal]{IEEEtran}

\usepackage{enumerate}
\usepackage{algorithm,algorithmic}
\usepackage{amssymb,amsfonts,amsmath}
\usepackage{float}
\usepackage{bbold}
\usepackage{cite}

\newcommand{\ones}{\boldsymbol{\mathbb{1}}}

\newtheorem{observation}{Observation}
\newtheorem{customproblem}{Problem}
\newenvironment{problem}[1]
  {\renewcommand\thecustomproblem{#1}\customproblem}
  {\endcustomproblem}

{\begin{list}               
    {$\bullet$ \hfill}{
        \setlength{\leftmargin}{\parindent}
        \setlength{\parsep}{0.04\baselineskip}
        \setlength{\itemsep}{0.5\parsep}
        \setlength{\labelwidth}{\leftmargin}
        \setlength{\labelsep}{0em}}
    }
{\end{list}}

\DeclareMathOperator*{\argmin}{arg\,min}
\providecommand{\cref}[1]{Chapter~\ref{chap:#1}}

\providecommand{\R}{\ensuremath{\mathbb{R}}}

\providecommand{\Z}{\ensuremath{\mathbb{Z}}}

\renewcommand{\vec}[1]{\ensuremath{\boldsymbol{#1}}}
\providecommand{\mat}[1]{\ensuremath{\boldsymbol{#1}}}
\providecommand{\wh}[1]{\ensuremath{\widehat{#1}}}
\providecommand{\wt}[1]{\ensuremath{\widetilde{#1}}}

\providecommand{\calS}{\mathcal{S}}
\providecommand{\calP}{\mathcal{P}}

\providecommand{\calM}{\mathcal{M}}

\providecommand{\calX}{\mathcal{X}}

\providecommand{\calC}{\mathcal{C}}
\providecommand{\calD}{\mathcal{D}}
\providecommand{\calS}{\mathcal{S}}
\providecommand{\calM}{\mathcal{M}}
\providecommand{\calW}{\mathcal{W}}

\providecommand{\mI}{\mat{I}}

 \providecommand{\vd}{\vec{d}}

\providecommand{\vn}{\vec{n}}  

 \providecommand{\vp}{\vec{p}}

\providecommand{\vx}{\vec{x}}

\providecommand{\vomg}{\vec{\omega}}

\ifCLASSINFOpdf
  \usepackage[pdftex]{graphicx}
\else
\fi
\usepackage{amsmath}

\begin{document}
%
\title{Super Resolution Phase Retrieval for Sparse Signals}
%
%
%

\author{Gilles~Baechler,~\IEEEmembership{Student~Member,~IEEE,}
        Miranda~Krekovi\'c,~\IEEEmembership{Student~Member,~IEEE,} Juri~Ranieri,\\
        Amina~Chebira,
        Yue M. Lu,~\IEEEmembership{Member,~IEEE,}
        and~Martin~Vetterli,~\IEEEmembership{Fellow,~IEEE}
\thanks{G. Baechler, M. Krekovi\'c and M. Vetterli are with EPFL, Switzerland, J. Ranieri is with Google, Switzerland, A. Chebira is with CSEM, Switzerland, and Y. M. Lu is with Harvard University, USA.}
\thanks{GB, MK, and JR had equal contributions and should be considered first authors. The work was divided as follows: JR, AC, YL and MV designed research, JR devised the support recovery algorithm and its performance bound, GB and MK proposed algorithmic improvements and carried out experiments, and GB, MK and JR wrote the manuscript.}
}

%
%

\markboth{IEEE Transactions on Signal Processing,~Vol.~XX, No.~X, July~2018}%
{Ranieri \MakeLowercase{\textit{et al.}}: Super Resolution Phase Retrieval for Sparse Signals}
%



\maketitle

\begin{abstract}

In a variety of fields, in particular those involving imaging and optics, we often measure signals whose phase is missing or has been irremediably distorted. Phase retrieval attempts to recover the phase information of a signal from the magnitude of its Fourier transform to enable the reconstruction of the original signal.
Solving the phase retrieval problem is equivalent to recovering a signal from its auto-correlation function.
In this paper, we assume the original signal to be sparse; this is a natural assumption in many applications, such as X-ray crystallography, speckle imaging and blind channel estimation. 
We propose an algorithm that resolves the phase retrieval problem in three stages: i) we leverage the finite rate of innovation sampling theory to super-resolve the auto-correlation function from a limited number of samples, ii) we design a greedy algorithm that identifies the locations of a sparse solution given the super-resolved auto-correlation function, iii) we recover the amplitudes of the atoms given their locations and the measured auto-correlation function. Unlike traditional approaches that recover a discrete approximation of the underlying signal, our algorithm estimates the signal on a continuous domain, which makes it the first of its kind. 

Along with the algorithm, we derive its performance bound with a theoretical analysis and propose a set of enhancements to improve its computational complexity and noise resilience. Finally, we demonstrate the benefits of the proposed method via a comparison against Charge Flipping, a notable algorithm in crystallography.


\end{abstract}


%
\IEEEpeerreviewmaketitle

\section{Introduction}

Imagine that instead of hearing a song you can only see the absolute value of its Fourier transform (FT) on a graphic equalizer. Can you recover the song from just this visual
information? The general answer is ``No'' as there exist infinitely many signals that fit the curve displayed by the equalizer. However, if we have additional information (or priors) about the song, we may be able to recover it successfully. The reconstruction process is the subject of this paper and is generally known as phase retrieval (PR).

Beside this day-to-day example, PR is of great interest for many real-world scenarios, where it is easier to measure the FT of a signal instead of the signal itself. During the measurement process, it may happen that the phase of the FT is lost or distorted.
Phase loss occurs in many scientific disciplines, particularly those involving optics and communications; a few examples follow.
\begin{itemize}
\item X-ray crystallography: we measure the diffraction pattern of a crystallized
  molecule---that is the magnitude of its FT---and we would like to recover
  the structure of the molecule itself~\cite{Millane:1990pt}.
\item Speckle imaging in astronomy: we measure many images of an astronomic
  subject and the phase of the images is compromised by the atmospheric
  distortion. We would like to recover the subject without the resolution
  downgrade imposed by the atmosphere~\cite{Knox:1976wl}.
\item Blind channel estimation of multi-path communication channels: we
  measure samples of the channel output without knowing the input. We would like to estimate the impulse response of the channel to optimize its capacity~\cite{Barbotin:2012vt}.
\end{itemize}

\subsection{Previous work}
The field of phase retrieval was born along with X-ray crystallography, when the first structures were solved with trial-and-error methods leveraging crystal symmetries. 
These initial attempts prepared the ground for more systematic approaches, a first example of which was proposed by Patterson in 1935~\cite{Patterson:1935wy}. This method is based on locating the peaks of the Patterson function---the auto-correlation function of the electron density---to determine pairwise differences between the locations of the atoms constituting a molecule. 

In the 1950s, a rich family of approaches exploiting the unique relationships between intensities and phases of measured diffraction patterns was developed, e.g. Cochran~\cite{Cochran:1955ch}, Sayre~\cite{Sayre:1952bc}, Karle~\cite{Karle:1950vh}.
These methods operate in the Fourier space and are known as \emph{direct methods} because they seek to solve the phase problem directly based on the observed intensities.

We would also like to emphasize the relevance of \emph{dual-space} algorithms, where both spatial and Fourier domains play a fundamental role in reconstructing the signal. While the origin of these methods dates back to 1972 with the work of Gerchberg and Saxton~\cite{Gerchberg:1972un}, a lot of interest was recently sparked by the introduction of \emph{Charge Flipping}~\cite{Oszlanyi:2004gb,Oszlanyi:2008fx}.

This short literature review of phase retrieval algorithms in X-ray crystallography is focused on \emph{ab initio} methods, that attempt to solve the phase problem with zero or very little prior information about the structure we are trying to infer. Hence, \emph{ab initio} methods are considered very challenging, given the minimal amount of information they have access to. Successful methods hinge on the design of an abstract data structure that reduces the degrees of freedom of the desired signal and simplifies its reconstruction. For example, direct methods exploit statistical relationships between the phases to reduce the number of unknowns, while Charge Flipping considers a discretization of the electron density.

\begin{figure*}[t!]
\centering
    \includegraphics[width=.98\linewidth]{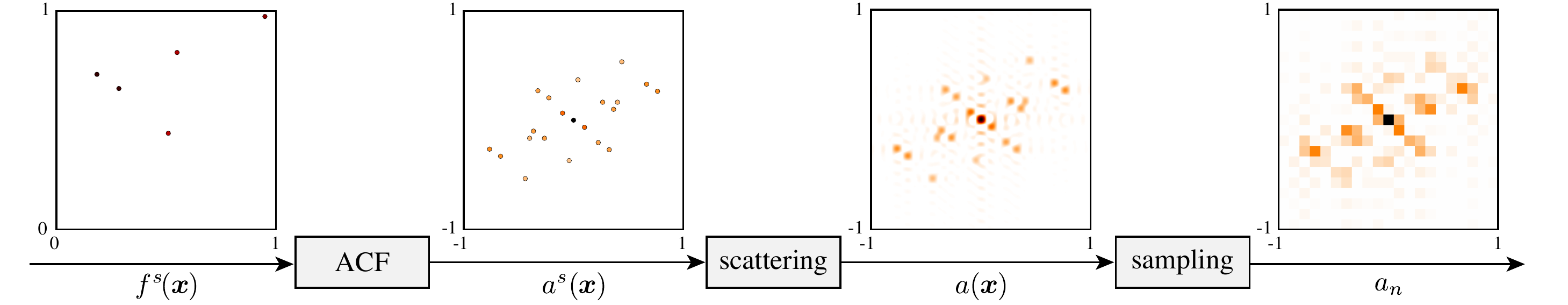}
    \caption{Typical PR measurement pipeline: the signal of interest  $f^s(\vx)$ generates the auto-correlation function $a^s(\vx)$, which is first filtered by the scattering function $\psi(\vx)$ (here an ideal lowpass filter) to yield $a(\vx)$ and then sampled, resulting in $a_{\vec{n}}$. Note that the spatial samples $a_{\vec{n}}$ can be obtained via the inverse discrete FT of the Fourier samples $A_{\vec{m}}$, when the periodicity in the two domains holds. Darker colors represent higher intensities.}
    \label{fig:sampling_scheme}
\end{figure*}

In this paper, we focus on the PR problem on sparse signals. The sparsity assumption is legitimate and encountered in many applications; for example atoms in crystallography form a sparse structure. We consider the most compact structure one can imagine for a sparse signal: a set of $K$ atoms defined by their locations $\vx_k$ and their amplitudes $c_k$,
\begin{align}
    f(\vx)=\sum_{k=1}^K c_k\phi(\vx-\vx_k)=f^s(\vx)*\phi(\vx),
    \label{eq:sparse_model}
\end{align}
where $f^s(\vx) = \sum_{k=1}^K c_k \delta(\vx - \vx_k)$ represents the structure, $\vx$ is a spatial variable defined over $\R^D$ (with $D$ being the
dimensionality of the signal), $\phi(\vx)$ is the scattering function induced by one atom and $*$ is the convolution operator.

Even if the advantage of the compact model defined in~\eqref{eq:sparse_model} looks appealing, the associated algorithmic challenges are often overwhelming. Computer scientists attempted to design a scalable (i.e. with a computational complexity that is polynomial in the number of atoms $K$) and stable to noise algorithm that could solve all possible instances of this problem without much success; to date, it is not even clear that such an algorithm would exist~\cite{Skiena:1994uo}.
In other words, we encounter a nontrivial trade-off  between the compactness of such data structures (i.e. the number of unknown variables) and the ease of solving the PR problem using them. For example, Charge Flipping easily solves many
PR problems in X-ray crystallography, but it is based on a discrete spatial structure, which is not the most compact representation.

Recently, we observed the emergence of new PR algorithms leveraging the notion of sparsity while assuming a discrete spatial domain. Two notable examples are \emph{GrEedy Sparse PhAse Retrieval} (GESPAR)~\cite{Shechtman2013}, based on the \emph{2-opt} algorithm~\cite{Croes195}, and \emph{Two-stage Sparse Phase Retrieval} (TSPR)~\cite{JaganathanOH13}, where the support is recovered by solving the discrete \emph{turnpike problem}~\cite{Dakic2000, Skiena1990}.
Both algorithms differ from our approach in that their models are discrete and the locations are bound to a discrete grid. Even though it was not designed with continuous setups in mind, TSPR can theoretically recover locations on a continuous domain. However, while it handles noise on the measured coefficients, it does not tolerate noise in the support, which makes it impractical for continuous setups.

The major benefit of having a continuous parametric model is that it enables estimation of the locations and amplitudes avoiding any discretization. In such a case, the achievable resolution is theoretically infinite and only limited by the noise corrupting the
measurements. This is what we call \emph{super resolution} phase retrieval.

\subsection{Main contributions and outline}

We propose a three-stage framework that precisely determines a sparse signal from the absolute value of its FT. In Section~\ref{sec:problem_statement},
we formalize the problem and describe the typical PR measurement pipeline. In Section~\ref{sec:three_stage}, we give a high-level overview of our modular approach, discuss the main challenges and introduce a few relevant properties.
Algorithms to solve these different modules are proposed in Section~\ref{sec:algorithms}. 

We then describe the details of the proposed method to recover the support, which constitutes the critical element of the pipeline: its complexity analysis can be found in Section~\ref{sec:complexity_analysis}, together with a method to reduce its computational cost, while
Section~\ref{sec:performance_analysis} identifies a theoretical bound (confirmed by numerical simulations) to successfully recover the signal support in a noisy regime. Then in Section~\ref{sec:improvements}, we propose a few improvements and variations of the algorithm to make it more robust to noise. In Section~\ref{sec:star}, we discuss the influence of the support configuration on the resulting reconstruction.
Finally, Section~\ref{sec:charge_flipping} compares our PR pipeline with the state-of-the-art.

Throughout this paper, we use bold lower case letters for vectors and bold upper case letters for matrices. Upper case calligraphic letters denote sets, e.g. $\calX$. Furthermore, $\wt{\calX}$ represents a set with noisy elements and $\wh{\calX}$ an estimated set.
Subscripts are reserved for indexing elements in lists and vectors.
In the primal domain, continuous functions are written in lower case letters and indexed with $\vx$, e.g. $f(\vx)$ and discrete functions are indexed with $\vn$, e.g. $f_{\vn}$. In the Fourier domain, we use capital letters, that is $F(\vomg)$ and $F_{\vec{m}}$, for continuous and discrete functions, respectively.




\section{Problem statement}
\label{sec:problem_statement}
We consider the FT of the signal defined in \eqref{eq:sparse_model},
\begin{align}
    F(\vomg)=\sum_{k=1}^K c_k\exp\left\{-j\vomg^\top \vx_k\right\}\Phi(\vomg), 
    \label{eq:FT}
\end{align}
where $\vomg$ is the frequency variable and $\Phi(\vomg)$ is the FT of the known kernel $\phi(\vx)$.

In practice, it is impossible to measure the whole FT \eqref{eq:FT}, hence we sample it. Furthermore, due to limitations of the measurement setup, we are usually only able to measure the absolute values of such samples, that we denote $|F_{\vec{m}}|$, where $F_{\vec{m}} = F(\vec{m}\Omega)$, $\vec{m} = \Z^D$ and $\Omega$ is the sampling frequency. As previously mentioned, the PR problem has infinite solutions without a priori knowledge of the signal $f(\vx)$, since we can assign any phase to the measurements and obtain a plausible reconstruction.
The role of structures, such as \eqref{eq:sparse_model}, is to constrain the PR problem to a correct and possibly unique solution. Under the sparsity assumption, retrieving the phase is equivalent to retrieving the locations and amplitudes of $f(\vx)$.

The auto-correlation function (ACF) $a(\vx)$ of $f(\vx)$ is given by the inverse FT of $|F(\vomg)|^2$:
\begin{align}
    a(\vx)=f(\vx)*f(-\vx) = \mathcal{F}^{-1}\left[|F(\vomg)|^2\right],
\end{align}
where $\mathcal{F}^{-1}$ is the inverse FT operator~\cite{vetterli2014foundations}.
Interestingly, the ACF structure is completely inherited from the signal \eqref{eq:sparse_model}:
\begin{align}
    a(\vx)&=\sum_{k=1}^K\sum_{\ell=1}^K c_k c_\ell \psi(\vx-(\vx_k-\vx_\ell)) \nonumber \\
    &=\left[\sum_{k=1}^K\sum_{\ell=1}^K c_k c_\ell \delta(\vx-(\vx_k-\vx_\ell))\right]*\psi(\vx) \nonumber \\
    &=a^s(\vx)*\psi(\vx),
    \label{eq:sparse_ACF}
\end{align}
where the kernel $\psi(\vx)$ is the ACF of $\phi(\vx)$ and $a^s(\vx)$ is the
ACF of the sparse structure of the train of Diracs $f^s(\vx)$. Equivalently, in the Fourier domain, we have
\begin{align}
\label{eq:acf_ft}
A(\vomg)=\sum_{k=1}^K \sum_{\ell=1}^K c_k c_\ell \exp\left\{-j\vomg^\top (\vx_k - \vx_\ell) \right\}|\Phi(\vomg)|^2.
\end{align}

The PR acquisition pipeline can be summarized as the filtering of the ACF $a^s(\vx)$ followed by sampling, where the filtering represents the scattering operation, as illustrated in Fig.~\ref{fig:sampling_scheme}.
We now have all the ingredients to state the core problem of this paper. 

\begin{problem}{1}
\label{problem:super_resolution_phase_retrieval}
Given Fourier samples $A_{\vec{m}} = A(\vec{m} \Omega)$ of the sparse ACF defined in~\eqref{eq:sparse_ACF}, recover the support $\calX=\{\vx_k\}_{k=1}^K$ and amplitudes $\{c_k\}_{k=1}^K$ determining the signal $f(\vx)$.
\end{problem}

Note that the information we are interested in is hidden behind two walls: the convolution with the kernel $\psi(\vx)$ that spatially blurs the sparse structure of the ACF and the phase loss of the original sparse signal, $f^s(\vx)$, that usually characterizes any PR problem.

\section{A three-stage approach}
\label{sec:three_stage}
We propose to solve Problem~\ref{problem:super_resolution_phase_retrieval} in three distinct stages: i) reconstruct the continuous ACF $a(\vx)$ from a set of its discrete Fourier coefficients, ii) estimate the support $\calX$ of $f(\vx)$ given $a(\vx)$, and iii) estimate its amplitudes $\{c_k\}_{k=1}^K$.


The first step is a classical sampling problem where we would like to fully characterize a continuous signal from a set of discrete measurements.

\begin{problem}{1.A}[Sparse ACF super resolution]
\label{problem:fri}
Given samples $A_{\vec{m}}$ of the sparse ACF as defined in~\eqref{eq:sparse_ACF}, recover its continuous version $A(\vec{\omega})$.
\end{problem}

The most well-known sampling result is due to Nyquist-Shannon-Kotelnikov and guarantees perfect recovery for signals that lie in the subspace of bandlimited functions, provided that the sampling rate is high enough. 

In our case, $f(\vx)$ and $a(\vx)$ are obviously not bandlimited, but we assume that such signals are sparse, as in~\eqref{eq:sparse_model}. Sparsity has two antagonistic effects on PR: it makes the problem combinatorial and hence hard to solve, but at the same time enables a divide-and-conquer approach, in which we first recover the support $\calX$ and then the amplitudes of $f(\vx)$. We argue that the support contains more information than the amplitudes, hence we choose to estimate it first.
As an example, if all the atoms have the same amplitude, then only the support is useful to recover the original signal. On the other hand, if all the atoms have the same location, the problem is trivially solvable.

\begin{problem}{1.B}[Support recovery]
Assume we are given the complete set of unlabeled differences $\calD=\{\vd_{k,\ell}\}_{k,\ell}=\{\vx_k-\vx_\ell\}_{k,\ell}$, recover the support $\cal{X}$ of the sparse signal $f(\vx)$.
\label{prb:PR}
\end{problem}

In most real-world scenarios, the unlabeled differences of Problem~\ref{prb:PR} are corrupted by noise. Hence, we assume additive Gaussian noise affecting $\vd_{k,\ell}$,
\begin{align}
\label{eq:noise}
\vec{\widetilde{d}}_{k,\ell} = \vec{d}_{k,\ell} + \vec{\nu}_{k, \ell},
\end{align}
where $\vec{\nu}_{k, \ell} \sim \mathcal{N} (\mathbf{0}, \sigma \mI)$. Furthermore, we denote the set of measured differences as $\wt{\calD}= \{ \vec{\widetilde{d}}_{k,\ell}\}_{k,\ell}$. For simplicity of notation, we convert the pairs of indices $(k, \ell)$ to $n \in \{1, \hdots, N\}$, where $N = K^2 - K + 1$, and order them such that $\| \vec{\widetilde{d}}_1 \| \leq \| \vec{\widetilde{d}}_2 \| \leq \hdots \leq \| \vec{\widetilde{d}}_N \|$. We do not assume any ordering on the elements of $\calX$.

In what follows, we state a few interesting observations related to Problem~\ref{prb:PR}.
First, when we measure a set of differences, some information is inevitably lost. 

\begin{observation}
\label{obs:shift_translate}
A set of points can be reconstructed from their pairwise differences, even when labeled, only up to shifts and reflections.
\end{observation}

To show that, we first translate and reflect the set of points $\calX$ as $\calX'=-\calX + \bar{\vx}$, where we overload the arithmetic operators on sets to transform each point as $\vx'_{k}=- \vx_k + \bar{\vx}$. Then, the set of differences of the transformed points is equivalent to the original one,
\begin{align}
    \vd'_{k, \ell}=\vx'_{k}-\vx'_{\ell}=-\vx_{k} + \bar{\vx} + \vx_{\ell} - \bar{\vx}=\vx_{\ell}-\vx_{k}=-\vd_{k,\ell},\nonumber
\end{align}
where the natural symmetry of $\calD$ compensates for the negative sign.

Second, while excluding shifts and reflections does not lead to a unique solution in general, we can still prove uniqueness under certain assumptions.

\begin{observation}
Assume that the points $\vx_k$ are drawn independently at random from a sufficiently smooth distribution, then the solution is unique \cite{Ranieri:2013tx}.
\end{observation}

Third, we briefly discuss the occurrence of \emph{collisions} in the ACF. We say that there is a collision in the ACF when two different pairs of distinct points from $\calX$ map to the same difference in $\calD$. Since we consider a continuous domain for the support, it natively prevents the appearance of collisions.\footnote{The support recovery algorithm proposed in this paper can in fact handle collisions and could be used on a discretized space as well. However, assuming no collisions simplifies the recovery of the amplitudes and enables a few improvements to make the algorithm more resilient to noise.}

\begin{observation}
\label{obs:collisions}
If the locations of the points are independently drawn uniformly from a finite interval, then collisions in the ACF occur with probability zero.
\end{observation}

Last, we note that the set of differences $\cal{D}$ contains many valid solutions. In particular, we can construct two solutions from every element of $\cal{X}$; this is a direct consequence of Observation~\ref{obs:shift_translate}.

\begin{observation}
\label{obs:superset}
The set of differences $\cal{D}$ is a superset of $2K$ valid solutions $\wh{\calX}$ to Problem~\ref{prb:PR} and such solutions always contain the point zero, that is $\mathbf{0} \in \wh{\calX}$.
\end{observation}


To verify this, we pick an element of the support, e.g. $\vx_\ell$, and build the following tentative solution,
\begin{align}
\label{eq:solution_form}
    \wh{\calX}=\{\vx_k - \vx_\ell \mid k = 1, \dots, K \}.
\end{align}
Then, we notice that i) $\wh{\calX}$ is a valid solution with the shift fixed as $-\vx_\ell$, ii) $\wh{\calX} \in \cal{D}$ and iii) we have a solution for every element of $\cal{X}$. Moreover, we can exploit the symmetry of the ACF to reach the aforementioned $2K$ solutions.
Such an observation can be extended to the noisy case, assuming we allow the solution to be noisy as well. This property is essential to the algorithm for support recovery proposed in the next section.


Once the support $\wh{\calX}$ of the solution has been retrieved, it remains to find the amplitudes $\{c_k \}_{k=1}^K$ of the signal $f(\vx)$.

\begin{problem}{1.C}[Amplitude recovery]
Given an ACF $a(\vx)$ as defined in~\eqref{eq:sparse_ACF} together with the estimated support $\wh{\calX}$ of $f(\vx)$, find the amplitudes $\{c_k \}_{k=1}^K$.
\label{prb:AR}
\end{problem}


\section{Algorithms}
\label{sec:algorithms}
In this section, we lay down our solutions to Problems~\ref{problem:fri},~\ref{prb:PR} and~\ref{prb:AR}, effectively providing an end-to-end framework to solve the sparse PR problem.

\subsection{ACF super resolution}
\label{sec:fri}
When we look at \eqref{eq:sparse_ACF}, we notice that $a(\vx)$ is completely defined by the locations $\vx_k -\vx_{\ell}$ and the amplitudes $c_{k} c_{\ell}$. Hence, we can recast Problem~\ref{problem:fri} as a parameter estimation problem given the measured samples $A_{\vec{m}}$ of the FT of the ACF. An effective existing approach is known as \emph{finite rate of innovation} (FRI) sampling~\cite{Vetterli2002, Dragotti2007}. FRI-based methods are inspired by spectral analysis techniques to estimate the locations $\vx_k -\vx_{\ell}$; in what follows, we review their fundamentals for the sake of completeness. In this subsection, we restrict ourselves to the $1$-dimensional case for clarity, even though our implementation is generalized to higher dimensions (see~\cite{Maravic2004} for more details).

The essential ingredient in FRI is to represent the signal of interest as a weighted sum of complex exponentials in the following form:
\begin{align}
\label{eq:FRI_signal}
b_m = \sum_{n=1}^N \alpha_n u_n^m.
\end{align}
This formulation has several similarities with~\eqref{eq:acf_ft}; to see this, we define $t_n = x_k-x_{\ell}$, substitute $\alpha_n = c_k c_{\ell}$ and $u_n = \exp(-j \Omega t_n )$ and rewrite the sampled ACF $A_m$ as follows,
\begin{align}
\label{eq:acf_dft}
A_m = \sum_{n=1}^N \alpha_n u_n^m |\Phi(m \Omega)|^2.
\end{align}
We remark that $|\Phi(m \Omega)|^2$ does not allow us to express \eqref{eq:acf_dft} as a sum of complex exponentials yet. However, if we assume that the kernel function $\phi(x)$ is an ideal low-pass filter\footnote{
The FRI theory has also been generalized to a wide range of kernels such as combinations of B-splines and E-splines~\cite{Dragotti2007, Unser:1999, Unser:2005} or even arbitrary sampling kernels~\cite{Uriguen2013}, where a linear operation enables to obtain the desired form~\eqref{eq:FRI_signal} from~\eqref{eq:acf_dft}.}, i.e. a sinc function, its FT becomes a box function. Thus, we can ignore such a kernel for some neighborhood of $m$ around zero, since $\Phi(m \Omega) = 1$ for $|m \Omega|$ smaller than the bandwidth of the signal.

The locations $\{ d_n \}_{n=1}^N$ are fully determined by the exponentials $\{ u_n \}_{n=1}^N$, that is $d_n = \frac{ \angle u_n }{\Omega}$, with $\angle u_n$ being the phase of $u_n$.
Recovering $u_n$ from~\eqref{eq:FRI_signal} is a classical spectral estimation technique and a possible solution is provided by \emph{Prony's method}~\cite{prony1795, stoica2005spectral}. The idea is to identify a filter $H_m$ to \emph{annihilate} $A_m$, which is mathematically defined as
\begin{align}
\label{eq:annihilating_eq}
(A * H)_m=0.
\end{align}
Then, the filter $H$ can be estimated by rewriting and solving~\eqref{eq:annihilating_eq} as a Toeplitz system.
As shown in~\cite{Vetterli2002}, if $A_m$ has the form of~\eqref{eq:FRI_signal}, then the $z$-transform of $H_m$ is 
\begin{align}
H(z) = \sum_{n=0}^N A_n z^{-n} = \prod_{n=1}^N (1- u_n z^{-1}),
\end{align}
where $u_n$ are nothing else but the roots of $H(z)$.
Our situation differs from usual FRI applications in the sense that the locations of the ACF describe a symmetric structure. As a consequence, all roots $u_n$ come in conjugate pairs (except for the one corresponding to the zero location).

Once the locations are known, the amplitudes $\alpha_n$ are found by injecting $u_n$ in~\eqref{eq:acf_dft} and solving a linear system of equations.

\subsection{Support recovery}
\label{sec:support_recovery}
For the recovery of the support, we propose a novel greedy algorithm that is initialized with a partial solution $\wh {\calX}_2$, which contains two locations. At a given iteration $k$, we generate a partial solution $\wh{\calX}_{k+1}$ composed of $k+1$ locations, hence the algorithm has a total of $K-2$ iterations indexed from $2$ to $K-1$.

\subsubsection{Initialization}

From Observation~\ref{obs:superset}, we know that the solution set $\wh{\calX}$ is contained in $\widetilde{\calD}$ and $\vec{0} \in \calX$; this gives us the first point of the solution, that is $\wh{\vx}_1 =\vec{0}$.
Next, we identify the element $\vec{\widetilde{d}}_N$ in $\widetilde{\calD}$ with the largest norm, so that we maximize the noise resilience of our algorithm. Indeed, assuming that the locations are corrupted by identically distributed noise, picking the largest norm ensures the maximal SNR of our initial solution. Note that the value $\vec{\widetilde{d}}_N$ is the noisy difference between two unknown locations of $f(\vx)$; without loss of generality, we call them $\vx_1$ and $\vx_2$. The elements $\wh{\vx}_1 = \vec{0}$ and $\wh{\vx}_2 = \vec{\widetilde{d}}_N$ are nothing but $\vx_1$ and $\vx_2 + \vec{\nu}_{2,1}$ translated by $-\vx_1$. Therefore, we are always guaranteed that the initialized solution $\widehat{\mathcal {X}}_2=\{\vec{0}, \vec{\widetilde{d}}_N \}$ is a (noisy) subset of the set of locations $\mathcal{X} - \vx_1$.

Reffering again to Observation~\ref{obs:superset}, we know that the set of differences $\widetilde{\cal{D}}$ contains the rest of the points $\{ \vx_k-\vx_1 + \vec{\nu}_{k,1} \}_{k=3}^{K}$, that should belong to the final solution $\wh{\calX} = \wh{\calX}_K$. Furthermore, since we do not want to duplicate points in $\wh{\calX}_k$, we initialize a set of possible elements of the solution $\mathcal{P}_2=\widetilde{\calD} \setminus  \{ \wt{\vd}_1, \wt{\vd}_N \}$. Due to noise, the vector $\vec{0}$ is not in $\wt{\calD}$, so we remove the closest element $\wt{\vd}_1$.

\begin{algorithm}[tb]
 \caption{Support recovery}
 \begin{algorithmic}
 \label{algo:support_recovery}
\REQUIRE A set of $N = K^2-K+1$ differences $\widetilde{\calD} = \{ \wt{\vd}_n \}_{n=1}^N$ ordered by their norms
\ENSURE A set of $K$ points $\wh{\calX}$ such that their pairwise differences generate $\widetilde{\calD}$
\STATE $\widehat{\calX}_2=\{\vec{0}, \wt{\vd}_N \}$
\STATE $\mathcal{P}_2=\widetilde{\calD} \setminus \{ \wt{\vd}_1, \wt{\vd}_N \}$
\FOR{$k = 2, \dots, K-1$}
    \STATE $\wh{\vx}_{k+1}=\argmin_{\vp\in\mathcal{P}_k}
    \sum_{\wh{\vx} \in \wh{\calX}_k} \min_{\vec{\widetilde{d}} \in \widetilde{\calD}} \left \| \vp-\wh
    {\vx}-\vec{\widetilde{d}} \right \|^2$
    \STATE $\wh{\calX}_{k+1} = \wh{\calX}_{k}\cup \wh{\vx}_{k+1}$
    \STATE $\mathcal{P}_{k+1}=\mathcal{P}_k\setminus \wh{\vx}_{k+1}$ 
\ENDFOR
\RETURN $\wh{\calX}_{K}$
 \end{algorithmic}
\end{algorithm}

\subsubsection{Main algorithm}

\begin{figure}[t!]
\centering
    \includegraphics[width=1\linewidth]{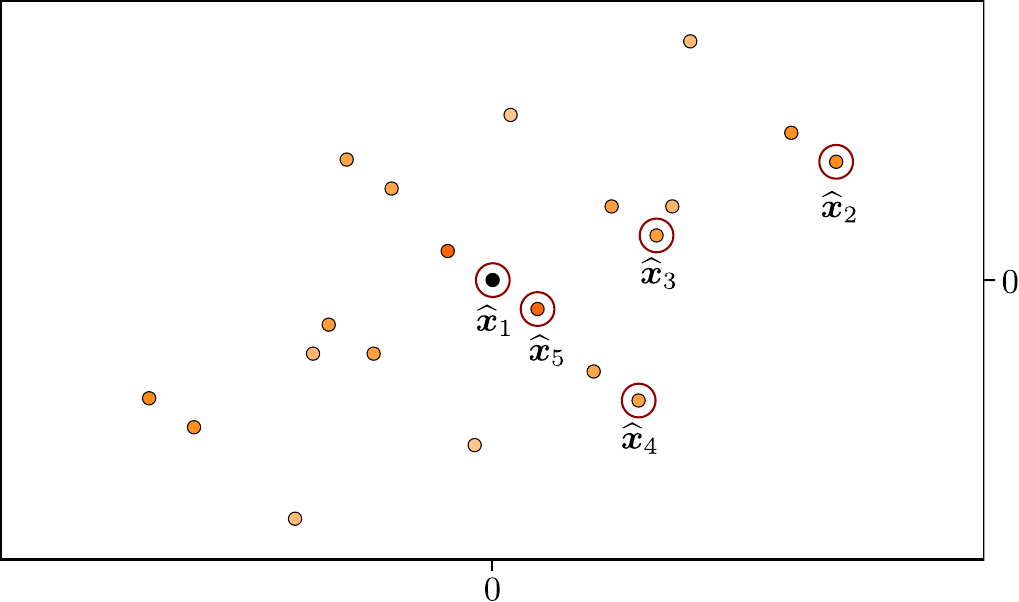}
    \caption{Instance of Algorithm~\ref{algo:support_recovery} on the ACF $a^s(\vx)$ from Fig.~\ref{fig:sampling_scheme}. We start by setting $\wh{\vx}_1 = \vec{0}$ and identifying $\wh{\vx}_2$, the point with the largest norm. Points $\wh{\vx}_3$ to $\wh{\vx}_5$ are then selected in a greedy way according to~\eqref{eq:cost_function}. The solution coincides with the initial signal $f^s(\vx)$ displayed in Fig.~\ref{fig:sampling_scheme}.}
    \label{fig:solution}
\end{figure}

At each step $k$, we identify the element in $\mathcal{P}_k$ that, when added to the partial solution $\wh{\calX}_{k}$, minimizes the error with respect to the measured set of differences $\widetilde{\cal{D}}$.
More precisely, at every iteration $k$ we solve the following optimization problem,
\begin{align}
    \wh{\vx}_{k+1}=\argmin_{\vec{p}\in\mathcal{P}_k}\sum_{\wh{\vx} \in \wh{\calX}_k} \min_{\wt{\vd} \in \widetilde{\calD}} \left \| \vec{p}-\wh
    {\vx}-\wt{\vd}\right \|^2.
   	\label{eq:cost_function}
\end{align}
Intuitively speaking, we would like to identify the element $\wh{\vx}_{k+1} \in \mathcal{P}_k$ such that the set of pairwise differences of the points in $\wh{\calX}_{k+1} = \wh{\calX}_{k} \cup \wh{\vx}_{k+1}$ is the closest to a subset of the measured $\wt{\calD}$. 
The main challenge is that we do not know the correct labeling between these two sets.
In the noiseless case, we are looking for a set $\wh{\calX}_{k+1}$, whose pairwise differences form exactly a subset of $\wt{\calD}$. Hence, we can solve the labeling by matching identical elements.
In the noisy case, we cannot leverage the definition of a subset. Therefore, we loosen the equality between elements and determine the labeling by searching for the differences in $\wt{\calD}$ that are closest in $\ell^2$-norm to the pairwise differences of the elements in $\wh{\calX}_{k+1}$.
This procedure is summarized in Algorithm~\ref{algo:support_recovery} and its application on the ACF $a^s(\vx)$ from Fig.~\ref{fig:sampling_scheme} is illustrated in Fig.~\ref{fig:solution}.

\subsection{Amplitude recovery}
\label{sec:amplitude_recovery}

If we assume that collisions can occur, recovering the amplitudes with a given ACF and support is equivalent to solving a system of quadratic equations. However, if there are no collisions, we suggest a simple but efficient algebraic solution to Problem~\ref{prb:AR}, inspired from~\cite{Ranieri:2013tx}. Our new approach is different in that it avoids a matrix inversion step and hence, it is both faster and more robust to noise.

Let $\vec{c} = [ c_1, c_2, \dots, c_K ]^\top$ be a vector made of the amplitudes to be recovered. If we define a matrix $\vec{C} = \vec{cc}^\top$, all the elements outside of the diagonal of such a matrix are the amplitudes of the measured ACF, that is $C_{i,j} = c_i c_j$. Notice that we cannot observe the diagonal entries $C_{i,i} = c_{i,i}^2$ as we just have access to their sum $a^s_{\vec{0}} = \sum_i c_{i,i}^2$, which is the value of the ACF at $\vec{0}$. This is unfortunate since they are precisely the values we are interested in, up to a squaring operator.

We recast Problem~\ref{prb:AR} as a matrix completion problem, where we would like to estimate the diagonal entries $C_{i,i}$ under the constraint of $\vec{C}$ being a rank-one matrix. 
The first step of our proposed method is to introduce a matrix $\vec{L}$ such that
\begin{align}
L_{i,j} = \begin{cases}
\log(C_{i,j}) = \ell_i + \ell_j & \mbox{for } i \neq j\\
0 & \mbox{otherwise},
\end{cases}
\end{align}
where $\ell_i = \log(c_i)$. The sum of the $i$th row of $\vec{L}$ is given by
\begin{align}
\sum_{j=1}^K L_{i,j} = (K-1) \ell_i + \sum_{j=1}^K \ell_j - \ell_i = (K-2) \ell_i + \sum_{j=1}^K \ell_j,
\end{align}
where the term $\sum_j \ell_j$ does not vary between rows. Hence, its value can be obtained from summing all the entries in $\vec{L}$,
\begin{align}
s = \sum_{i=1}^K \sum_{j=1}^K L_{i,j} &= (K-2) \sum_{i=1}^K \ell_i + K \sum_{j=1}^K \ell_j \nonumber \\ &= 2(K-1)\sum_{j=1}^K \ell_j.
\end{align}
Then, we recover the vector $\vec{\ell} = [ \ell_1, \ell_2, \dots, \ell_K ]^\top$ for $K > 2$ as
\begin{align}
\vec{\ell} = \frac{1}{K-2}\left(\sum_{j=1}^K L_{i,j} - \frac{s}{2(K-1)} \ones \right),
\end{align}
where $\ones$ is the all-ones vector.\footnote{When $K=2$, the entries $\ell_1$, $\ell_2$ can be recovered by solving a system of two equations.} Finally, it suffices to compute $c_i = \exp(\ell_i)$ to retrieve the amplitudes.

Note that this solution assumes that $\vec{C}$ is symmetric; this might not be the case in a noisy setup, but we enforce it by replacing $\vec{C}$ with $\frac{1}{2}(\vec{C} + \vec{C}^\top)$. In case of collisions, the problem does not have an algebraic solution and a possible convex relaxation is provided in~\cite{JaganathanOH13}. In practice, this is often not a concern due to Observation~\ref{obs:collisions}.

In what follows, we study and propose improvements to the performance of our PR algorithm, focusing our attention on the support recovery step, i.e. Algorithm~\ref{algo:support_recovery}. In fact, the first step---the super-resolution with FRI---is well represented in literature, where theoretical analyses, extensive simulations in noisy scenarios and efficient denoising schemes have been proposed~\cite{Maravic2004, Dragotti2009, Pan2017}. On the other hand, the amplitude recovery, while being novel, only consists of simple algebraic manipulations that are not computationally costly.

\section{Complexity analysis}
\label{sec:complexity_analysis}

\begin{figure}[tb]
\centering
    \includegraphics[width=1\linewidth]{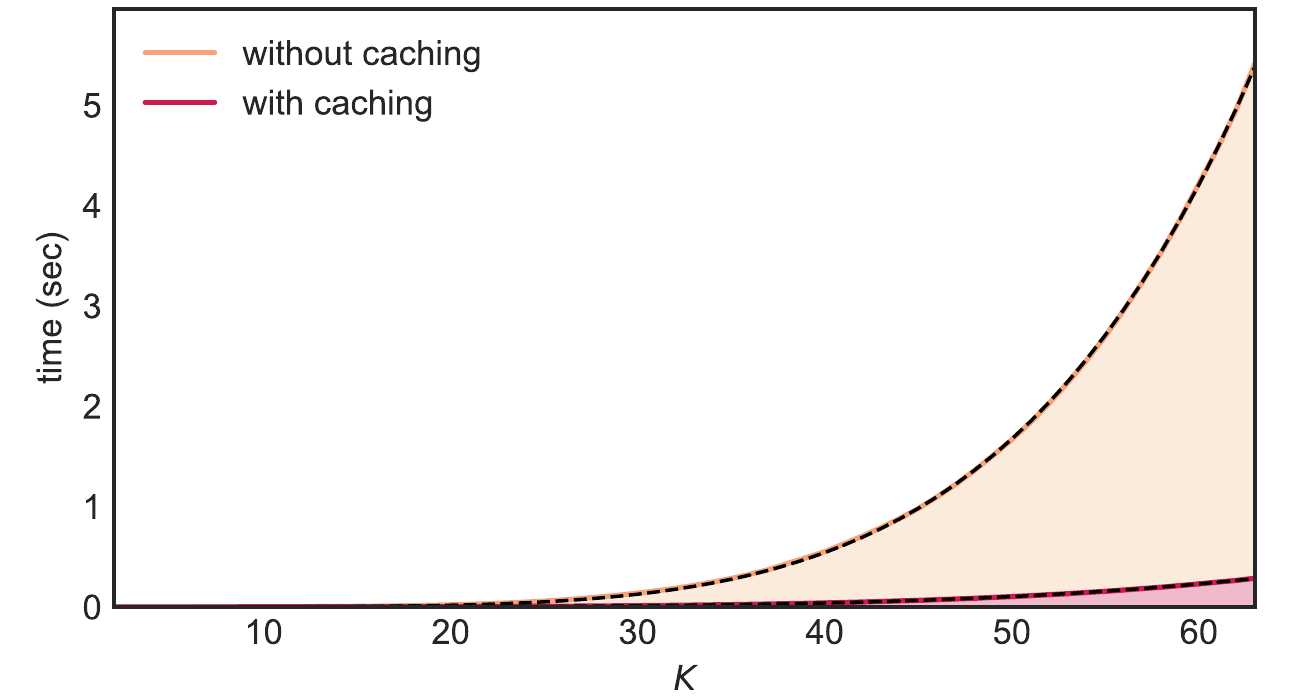}
    \caption{Comparison of the average run time of the original algorithm and its cached version. The times reported are the average of $100$ runs of the algorithm.
    The dashed lines represent curves of the form $C K^\alpha$ that are fitted to the data. For the method without caching, we have $C = 4.25 \cdot 10^{-6}$ and $\alpha = 5.06$, while for the method with caching we have $C = 3.88 \cdot 10^{-6}$ and $\alpha = 4.37$. Remark how the caching is reducing the polynomial degree of the computational cost by approximately one.}
    \label{fig:caching}
\end{figure}

Algorithm~\ref{algo:support_recovery} has $K$ rounds. In each of these rounds, we go through all points in the existing solution set $\wh{\calX}_k$, and for each point we compute the difference with all the values in $\widetilde{\calD}$. Since there are $\mathcal{O}(K)$ points in $\wh{\calX}_k$ and $\mathcal{O}(K^2)$ elements in $\widetilde{\calD}$, this is done in $\mathcal{O}(K^3)$ operations.
Furthermore, for each of these computed differences, we need to find the closest element in $\widetilde{\calD}$, which requires additional $\mathcal{O}(K^2)$ comparisons.
In total, the complexity of our algorithm is $\mathcal{O}(K^{6})$. Even though this is high and limits the field of application to reasonable sizes, it compares favorably to an exhaustive search strategy, which grows exponentially with $K$.

It is possible to trade time complexity for storage complexity. Indeed, we observe that we compute at each round the following values
\begin{align}
\label{eq:cached_values}
\wt{\vd}_{i,j} =  \argmin_{\vec{\widetilde{d}} \in \widetilde{\calD}} \| \vp_j -  \wh{\vx}_i- \vec{\widetilde{d}} \|^2,
\end{align}
for every point $\wh{\vx}_{i} \in \wh{\calX}_k$ and candidate $\vp_j \in {\calP}_k$. However, since we are just moving one element from ${\calP}_k$ to $\wh{\calX}_{k+1}$ at each iteration, we propose to cache the values~\eqref{eq:cached_values} in a lookup table to reduce the total computational cost. By doing so, we only need to update each $\wt{\vd}_{i,j}$ when the corresponding candidate $\vp_j$ is removed from ${\calP}_k$ to be added to $\wh{\calX}_{k+1}$.

The theoretical complexity when caching $\wt{\vd}_{i,j}$ is not trivial to analyze, but in practice we notice a significant improvement, as illustrated in Fig.~\ref{fig:caching}.

\section{Performance analysis}
\label{sec:performance_analysis}

In what follows, we study the expected performance of Algorithm~\ref{algo:support_recovery} in the presence of noise. More precisely, we estimate the expected mean squared error (MSE) of the support recovery algorithm when the correct solution is obtained in Section~\ref{sec:exp_perf}. Then, we approximate the probability of the algorithm to find such a correct solution in Section~\ref{sec:probability_of_success}. We consider the problem with $D=1$ dimension to lighten notation and simplify the discussion. However, all the results can be easily generalized to the multidimensional setup introduced in Problem~\ref{prb:PR}.

\subsection{Expected performance}
\label{sec:exp_perf}

After $K-2$ iterations and if the algorithm successfully finds any correct solution as defined in~\eqref{eq:solution_form}, then this solution will be noisy as it is constructed by selecting noisy elements from $\wt{\calD}$, see~\eqref{eq:noise}. 
If we assume Gaussian noise affecting the support, then the MSE of the support recovery solution can be computed as
\begin{align}
    \operatorname{MSE}=\|\calX-\widehat{\mathcal
    {X}}\|_2^2=\sum_{k=2}^{K}\|\vec{\nu}_{k,1}\|^2=\sigma^2 Q_{K-1},
\end{align}
where $Q_{K} \sim \chi^2_{K}$ follows a chi-squared distribution with $K$ degrees of freedom. Therefore, the expected value of the MSE of any correct solution is 
\begin{align}
\label{eq:expected_mse}
    \mathbb{E}[\operatorname {MSE}]=(K-1)\sigma^2.
\end{align}

\subsection{Probability of success}
\label{sec:probability_of_success}
We model the probability that Algorithm~\ref{algo:support_recovery} finds the correct solution as a function of the noise variance $\sigma^2$ and the number of elements $K$ to characterize its performance. 
Such a probability can be factored as $K-2$ iterations
\begin{align}
    P(\sigma,K)=\prod_{k=2}^{K-1}P_k(\sigma,K),
\end{align}
where $P(\sigma,K)$ is the probability of success of the support recovery algorithm and $P_k(\sigma,K)$ is the conditional probability of success at iteration $k$, given that the algorithm was correct until iteration $k-1$.

We focus our attention on what happens at iteration $k$, i.e. we study the probability $P_k(\sigma,K)$. First, we split the set of possible elements of the solutions $\mathcal{P}_k$ as the union of two disjoint sets: $\calC_k$ containing the elements that when picked by the algorithm generate a correct partial solution at iteration $k$, and $\calW$ containing the elements that when picked corrupt the partial solution. Second, we generalize the cost function used in the main optimization problem \eqref{eq:cost_function} to a generic set of $1$D elements $\mathcal{A}$ as,
\begin{align}
    g(\mathcal{A}, \wh{\calX}_k) = \min_{p\in\mathcal{A}}\sum_{\wh{x} \in \wh{\calX}_k} \min_{\wt{d} \in \widetilde{\calD}} \left( p-\wh{x}-\wt{d}\right)^2.
    \label{eq:generic_cost_function}
\end{align}
Below, we use $g(\mathcal{A}, \wh{\calX}_k)$ with both sets and single elements as arguments: in other words, the expression $g(a, \wh{\calX}_k)$ is interpreted as $g(\{a\}, \wh{\calX}_k)$.

Then, we compute the probability that the support recovery algorithm picks an element from $\calC_k$ instead of an element from $\calW$, when searching for the solution of \eqref{eq:cost_function}. This happens if the cost of $\calC_k$ is smaller than the one of $\calW$ measured via~\eqref{eq:generic_cost_function},
\begin{align}
\label{eq:p_k_sigma_K}
    P_k(\sigma,K)&=\mathrm{P}(g(\calC_k, \wh{\calX}_k) < g(\calW, \wh{\calX}_k)) \nonumber \\
    &=\mathrm{P}(\exists c\in\calC_k | g(c, \wh{\calX}_k) < g(\calW, \wh{\calX}_k))\nonumber \\
    &= 1- \mathrm{P}(\nexists c\in\calC_k\; |\; g(c, \wh{\calX}_k) < g(\calW, \wh{\calX}_k)) \nonumber \\
    &= 1- \mathrm{P}(\forall c\in\calC_k\; |\; g(c, \wh{\calX}_k) \geq g(\calW, \wh{\calX}_k)).
\end{align}
We assume that the events $g(c, \wh{\calX}_k) \geq g(\calW, \wh{\calX}_k)$ are independent for all $c \in \calC_k$ and obtain
\begin{align}
\label{eq:p_k_sigma_K_2}
    P_k(\sigma,K)&=1-\prod_{c\in\calC_k}\mathrm{P}\left(\frac{g(c, \wh{\calX}_k)}{g(\calW, \wh{\calX}_k)} \geq 1 \right).
\end{align}
This is a crude simplification, but it enables us to compute an approximation of $P_k(\sigma, K)$ that will not impair the quality of the end result, as we will demonstrate later.
With a similar development as~\eqref{eq:p_k_sigma_K}, we can write
\begin{align}
 \mathrm{P}\left(\frac{g(c, \wh{\calX}_k)}{g(\calW, \wh{\calX}_k)} \geq 1 \right) = 1- \mathrm{P}\left(\forall w\in\calW\; \bigm|\; \frac{g(c, \wh{\calX}_k)}{g(w, \wh{\calX}_k)} < 1 \right). \nonumber
\end{align}
Again, we approximate $P_k(\sigma,K)$ assuming the independence of the events $g(w, \wh{\calX}_k)$ as
\begin{align}
    P_k(\sigma,K)&= 1-\prod_{c\in\calC_k} \left( 1 - \prod_{w\in\calW}\mathrm{P}\left(\frac{g(c, \wh{\calX}_k)}{g(w, \wh{\calX}_k)} < 1\right) \right).
    \label{eq:prob_dev1}
\end{align}

Then, we focus our attention on the term $\mathrm{P}\left(\frac{g(c, \wh{\calX}_k)}{g(w, \wh{\calX}_k)} < 1 \right)$. First, we compute the cost of adding an element $c$ from $\calC_k$ to $\wh{\calX}_{k+1}$,
\begin{align}
\label{eq:prob_dev5_}
    g(c, \wh{\calX}_k)& = \sum_{\wh{x} \in \wh{\calX}_k} \min_{\wt{d} \in \widetilde{\calD}}
    \left(c-\wh{x}-\wt{d}\right)^2 \nonumber \\
    &=\sum_{\ell=1}^{k} \min_{\wt{d} \in \widetilde{\calD}}
    \left(c-(x_\ell-x_1+\nu_{\ell,1})-\wt{d}\right)^2, 
\end{align}
where each $\wh{x} \in \wh{\calX}_k$ is a shifted noisy version of an element of $\calX$. Following a similar reasoning, the newly added element $c$ can be expressed as $c = x_{k+1}-x_1+\nu_{k+1,1}$. By substituting this expression into~\eqref{eq:prob_dev5_}, we further obtain
\begin{align}
    g(c, \wh{\calX}_k)
    &=\sum_{\ell=1}^{k} \min_{\wt{d} \in \widetilde{\calD}}
    \left(x_{k+1}+\nu_{k+1,1}-x_\ell-\nu_{\ell,1}-\wt{d}\right)^2 \nonumber \\
    &\stackrel{(a)}{\approx}\sum_{\ell=1}^{k}
    \left(\nu_{k+1,1}-\nu_{\ell,1}-\nu_{k+1,\ell}\right)^2
    \nonumber \\
    &=3\sigma^2Q^{(1)}_{k},
    \label{eq:notsurewhat}
\end{align}
where $Q^{(1)}_{k} \sim \chi^2_{k}$, and in $(a)$ we approximate $g(c, \wh{\calX}_k)$ by selecting the difference $\wt{d}=x_{k+1}-x_\ell+\nu_{k+1,\ell}$. We select this specific $\wt{d}$ as it is likely to be picked, provided that the noise variance $\sigma^2$ is small compared to the values $x_i$. This also significantly simplifies~\eqref{eq:notsurewhat} by dropping the random variables $x_1$, $x_{k+1}$, and $x_{\ell}$.

Second, we analyze the cost of making an error $g(w, \wh{\calX}_k)$ at iteration $k$---that is selecting any element $w \in \calW$ given $\wh{\calX}_k$:
\begin{align}
 g(w, \wh{\calX}_k)=\sum_{\wh{x} \in \wh{\calX}_k} \min_{\wt{d}\in \wt{\calD}}
 \left(w-\wh {x}-\wt{d}\right)^2.
 \label{eq:prob_dev2}
\end{align}
We express the minimum in~\eqref{eq:prob_dev2} as an exhaustive check of all the possible selections of $k$ differences from $\wt{\calD}$. To do so, we define $\calM_k$ as the set containing all the $k$-permutations of $\wt{\calD}$, and rewrite the probability of selecting a correct location $c$ instead of a wrong one $w$ for any given $c$ and $w$ from \eqref{eq:prob_dev1} as follows,
\begin{align}
 \mathrm{P}\left(\frac{g(c, \wh{\calX}_k)}{g(w, \wh{\calX}_k)} < 1 \right) &= \mathrm{P}\left(\frac{g(c, \wh{\calX}_k)}{e(w, \vec{\pi}, \wh{\calX}_k)} < 1, \; \forall \vec{\pi} \in \calM_k \right). \nonumber
\end{align}
Here, we introduced $e(w, \vec{\pi}, \wh{\calX}_k)$ as the cost for a given permutation $\vec{\pi}$,
\begin{align}
	e(w, \vec{\pi}, \wh{\calX}_k)=\sum_{i = 1}^k    \left(w-\wh{x}_i-\pi_i\right)^2,
    \label{eq:prob_dev3}
\end{align}
where the elements in $\vec{\pi}$ and $\wh{\calX}_k$ are indexed with $i$.

Once more, we assume that all these selections are independent to obtain
\begin{align}
\label{eq:prob_dev4}
 \mathrm{P}\left(\frac{g(c, \wh{\calX}_k)}{g(w, \wh{\calX}_k)} < 1 \right) &= \prod_{\vec{\pi} \in \calM_k}\mathrm{P}\left(\frac{g(c, \wh{\calX}_k)}{e(w,\vec{\pi},\wh{\calX}_k)} <
    1\right).
\end{align}

Finally, we discuss the probabilistic aspects of~\eqref{eq:prob_dev3}.
The terms $\omega$, $\wh{x}_i$ and $\pi_i$ are each made of the difference between two points plus a noise value. Indeed, they have the form
\begin{align}
p = x_i - x_j + \nu_{i,j}, \nonumber
\end{align}
for some specific indices $i$ and $j$. Assuming that the points in $\calX$ are uniformly distributed between $-0.5$ and $0.5$, and the noise is Gaussian with zero mean and variance $\sigma^2$, we can approximate~\eqref{eq:prob_dev3} as
\begin{align}
    e(w, \vec{\pi}, \wh{\calX}_k)
    &\stackrel{(a)}{\approx}\sum_{\ell=1}^k \left(\sum_{i=1}^6 Y_i + \sum_{j=1}^3
    Z_j
    \right)^2 \nonumber \\
    &\stackrel{(b)}{\approx}\sum_{\ell=1}^k \left( W + \sum_{j=1}^3 Z_j
    \right)^2 \nonumber \\
    &=\left(3\sigma^2+\frac{1}{2}\right)Q^{(2)}_{k},
    \label{eq:prob_dev5}
\end{align}
where $Q^{(2)}_{k} \sim \chi^2_{k}$, $Y_i\sim U[-0.5,0.5]$, $Z_j \sim\mathcal{N}(0,\sigma^2)$ and $W\sim\mathcal
{N}(0,\frac{1}{2})$. In $(a)$, we approximate the sum
by assuming independence between all the random variables and in $(b)$ we approximate the sum of six random variables uniformly distributed on $[-0.5,0.5]$ with a normal random variable with variance $\sigma^2 = \frac{1}{2}$.

We now have all the ingredients to compute the probability of success at iteration $k$~\eqref{eq:prob_dev1}, as 
\begingroup
    \fontsize{9.2pt}{12pt}\selectfont
\begin{align}
    \label{eq:final_bound}
    P_k(\sigma, K) &= 1-\prod_{c\in\calC_k}\left(1- \prod_{w\in\mathcal
    {W}}\mathrm{P}\left(\frac{g(c, \wh{\calX}_k)}{g(w, \wh{\calX}_k)} < 1\right)\right) \nonumber \\
    &\approx 1-\prod_{c\in\calC_k}\left( 1 - \prod_{w\in\mathcal
    {W}} \prod_{\calS \in \calM_k}\mathrm{P}\left(\frac{Q^{(1)}_{k}}{Q^{(2)}_{k}} <
    \frac {3\sigma^2+\frac{1}{2}}
    {3\sigma^2}\right) \right) \nonumber \\
    &= 1-\left( 1 - \mathrm{P}\left( \frac{Q^{(1)}_{k}}{Q^{(2)}_{k}} <
    \frac {3\sigma^2+\frac{1}{2}}
    {3\sigma^2}\right)^{|\calM_k||\calW|}\right)^{|\calC_k|} \nonumber\\
    &=1-\left( 1 - \mathrm{F}\left(\frac
    {3\sigma^2+\frac{1}{2}}{3\sigma^2},k,k\right)^{|\calM_k||\calW|}\right)^{|\calC_k|},
\end{align}
\endgroup
where $\mathrm{F}(x,k_1,k_2)$ is the cumulative distribution function of an
F-distribution with parameters $k_1$ and $k_2$; it can be calculated using regularized incomplete beta functions. Last, we determine the size of the sets as
\begin{align}
    |\calC_k|&= K-k, \nonumber \\
    |\calW|&=N-K = K^2-2K+1, \nonumber \\
    |\calM_k|&= N^k.
\end{align}
Note that as the number of points $K$ increases, these exponents grow faster and push any probability that is not $1$ to $0$; hence, we expect a steep phase transition.

Along the path of our analysis, we made a few rough assumptions that we cannot theoretically justify regarding the independence of events, e.g. in~\eqref{eq:p_k_sigma_K_2},~\eqref{eq:prob_dev1} and~\eqref{eq:prob_dev4}. While we would like to be more rigorous, we provide below numerical evidence that such assumptions hold in practice as the algorithm's performance exhibits a phase transition matching closely the derived theoretical bound~\eqref{eq:final_bound}.

\begin{figure}[t!]
\centering
    \includegraphics[width=\linewidth]{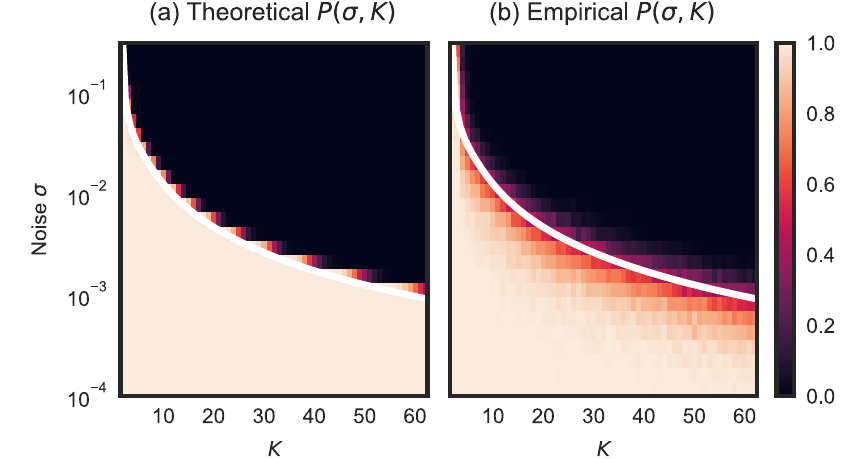}
    \caption{Comparison of the (a) theoretical and (b) empirical probability of success for Algorithm~\ref{algo:support_recovery} in 2 dimensions with respect to the size of the problem $K$ and the noise $\sigma$ affecting the set of differences.
    In both plots, the white line represents $P(\sigma, K) = 0.5$.}
    \label{fig:phase_transition}
\end{figure}

\subsection{Numerical simulations}
\label{sec:numerical_simulations}

We define the \emph{index-based} error as a binary metric that is equal to $0$ if the solution set $\wh{\calX}$ is of the form~\eqref{eq:solution_form}, and $1$ otherwise.
This error can be used to empirically measure the probability of success of Algorithm~\ref{algo:support_recovery}: we approximate it by running several experiments with different levels of noise $\sigma$ and numbers of points $K$.
In Fig.~\ref{fig:phase_transition}, we report the results of such an experiment and compare it with our theoretical result obtained in~\eqref{eq:final_bound}.
We confirm that $P(\sigma,K)$ exhibits a sharp phase transition---we can identify pairs $(K, \sigma)$ for which the algorithm always succeeds and pairs for which it always fails. However, the empirical phase transition is less sharp than the theoretical one and this is probably due to our approximations regarding the independence of events. Nonetheless, the two phase transitions are closely aligned for each value of $K$.

In the following, we develop some intuition that may explain why these approximations appear to be so tight. 
We claim that, even though not \emph{all} events are pairwise independent, \emph{most of them} are. As an example, when we look at 
\begin{align}
g(p, \wh{\calX}_k) = \sum_{\wh{x} \in \wh{\calX}_k} \min_{\wt{d} \in \widetilde{\calD}} \left( p-\wh{x}-\wt{d}\right)^2,
\end{align}
for two different values $p_1$ and $p_2$ of $p$, the respective cost functions $g(p_1, \wh{\calX}_k)$ and $g(p_2, \wh{\calX}_k)$ probably share a few common differences $\wt{d}$. However, we observe that at round $k$, only $k$ out of $K^2 -2K+1$ differences are selected, one for every $\wh{x} \in \wh{\calX}_k$. Then, assuming that most pairs $\left( g(p_1, \wh{\calX}_k), g(p_2, \wh{\calX}_k) \right)$ are independent is practically equivalent to assume that we select the differences $\wt{d}$ uniformly at random within the minimization. Moreover, we believe that the few dependent events ignored by such assumptions are one of the likely causes of the different steepness exhibited by the theoretical and observed phase transition.

\section{Improving noise resilience}
\label{sec:improvements}

We now discuss strategies and variations of our support recovery algorithm aiming at improving the quality of the solution in noisy settings. We chose not to include them in the analysis as they make it more intricate.

\subsection{Deleting solutions from the set of differences}

When a new point $\wh{\vx}_{k+1}$ is added to $\wh{\calX}_k$, Algorithm~\ref{algo:support_recovery} ignores some useful information. Assuming that there are no collisions and no noise, we know that the values $\wh{\calX}_k-\wh{\vx}_{k+1}$ and $\wh{\vx}_{k+1} - \wh{\calX}_k$ in $\calD$ cannot belong to the solution $\wh{\calX}$ as they are part of $\calW$. Thus, as soon as $\wh{\vx}_{k+1}$ is added to the solution set, we can remove all values of the form $\wh{\calX}_k - \wh{\vx}_{k+1}$ and $\wh{\vx}_{k+1} - \wh{\calX}_k$ from $\calD$.

The same reasoning applies to the noisy case, but we pick the closest values in $\wt{\calD}$ as we do not have exact differences. More formally, when we add a new point $\wh{\vx}_{k+1}$ to the solution $\wh{\calX}_k$, we dispose of the following $2k$ elements of $\wt{\calD}$,
\begin{align}
\wt{\vd}^* = \argmin_{\wt{\vd} \in \wt{\calD}} \| \pm \wh{\vx} \mp \wh{\vx}_{k+1} - \wt{\vd} \| ^2, \quad \forall \wh{\vx} \in \wh{\calX}_k. \nonumber
\end{align}
This approach results in two opposing effects. On one hand, we introduce the risk of erroneously discarding a point $\wt{\vd}^*$ that belongs to the solution. On the other hand, we are pruning many elements out of $\wt{\calD}$ and naturally reduce the risk of picking an erroneous candidate later on in the recovery process. As we will show in Section \ref{sec:comparison_improvements}, the benefits out-weight the risks.

\subsection{Symmetric cost function}

Next, we replace the cost function~\eqref{eq:cost_function} with a symmetric one to leverage the natural symmetry of the ACF.

In Algorithm~\ref{algo:support_recovery}, we search for the vectors in $\widetilde{\calD}$ closest to the computed differences $\vp - \wh{\calX}_k$ for each candidate $\vp$. We strengthen its noise resilience by jointly searching for the vectors closest to $\mp \wh{\calX}_k \pm \vp$ and choosing the candidate $\vp$ that minimizes the sum of both errors. Specifically, we rewrite the cost function~\eqref{eq:cost_function} as
\begingroup
    \fontsize{10pt}{12pt}\selectfont
\begin{equation}
\label{eq:cost_function_2}
    \wh{\vx}_{k+1}=\argmin_{\vp\in\mathcal{P}_k}
    \sum_{\wh{\vx} \in \wh{\calX}_k} \min_{\wt{\vd}, \wt{\vd}'\in \widetilde{\calD}} \left\|\vp - \wh{\vx}- \wt{\vd} \right \|^2 + \left \| \wh{\vx}-\vp- \wt{\vd}' \right \|^2. 
\end{equation}
\endgroup

We stress that this improvement is compatible with the idea of caching introduced in Section~\ref{sec:complexity_analysis}. We can in fact cache the following pairs
\begin{align}
(\wt{\vd}, \wt{\vd}')_{i, j} =  \argmin_{\wt{\vd}, \wt{\vd}'\in \widetilde{\calD}} \|\vp_j -\wh{\vx}_i-\wt{\vd} \|^2 + \| \wh{\vx}_i-\vp_j -\wt{\vd}'\|^2, 
\end{align}
for each $\wh{\vx}_i\in\wh{\calX}_k$ and $\vp_j \in \calP_k$
and recompute them when $\vp_j$ gets added to the solution $\wh{\calX}_{k+1}$.

\subsection{Denoising of partial solutions}

At each iteration $k$ of Algorithm~\ref{algo:support_recovery}, we have a partial solution $\wh{\calX}_{k+1}$ and, from~\eqref{eq:cost_function}, we identify for each pair $\wh{\vx}_i, \wh{\vx}_j \in \wh{\calX}_{k+1}$ a difference $\wh{\vd}_{i,j}$ that is the closest to $\wh{\vx}_i - \wh{\vx}_j$.
In other words, we are simultaneously labeling the differences $\wh{\vd}_{i,j}$ using our current partial solution; such a labeling is completed as $k$ reaches the final iteration $K-1$.

This partial labeling can be exploited to \emph{denoise} the set $\wh{\calX}_{k+1}$ as it provides unused additional constraints and mitigates the error propagation between the iterations. More precisely, we propose to find a set of points $\{ \wh{\vx}_i \}_{i=1}^{k+1}$ that minimizes the following cost function
\begin{align}
\label{eq:cdm_problem}
    J \big(\{ \wh{\vx}_i \}_{i=1}^{k+1} \big) = \sum_{i,j} \| \vec{\widehat{d}}_{i,j} - ( \wh{\vx}_i - \wh{\vx}_j) \|^2.
\end{align}
The solution to~\eqref{eq:cdm_problem} is derived in closed-form by setting its first derivative to $0$. As it is based on a least-square-error criterion, it is then optimal when the differences are corrupted by additive Gaussian noise.

\begin{figure}[tb]
\centering
    \includegraphics[width=1\linewidth]{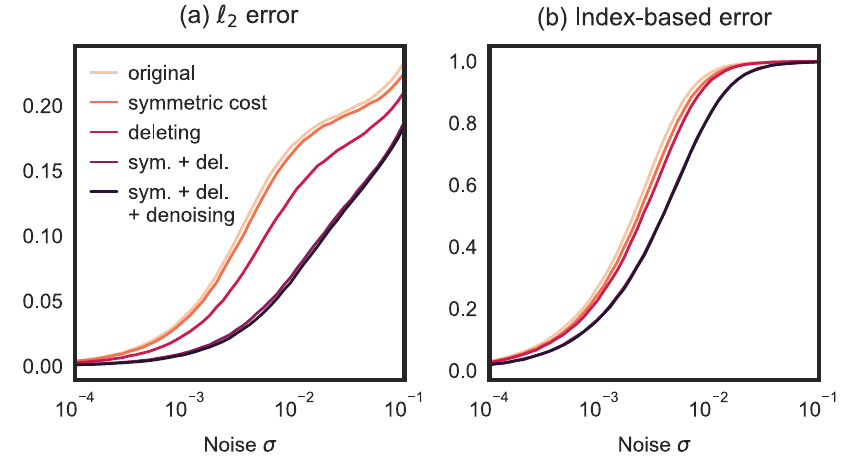}
    \caption{Average error for the different combinations of improvements of the algorithm. We create $\calX$ from $K=6$ $1$D points chosen uniformly at random from the interval $[0,1]$, create $\calD$ accordingly and add Gaussian noise $\mathcal{N}(0, \sigma^2)$ to its elements. The $\ell^2$ and the index-based errors are computed for different levels of noise $\sigma$ and different improvements of the original algorithm.  \label{fig:denoising_comparison}}
\end{figure}

This leads to a simple and effective strategy: refine the estimate of the solution set at each step by taking the average of the differences related to each point $\wh{\vx}_i\in\wh{\calX}_{k+1}$ as
\begin{align}
\label{eq:cdm}
\wh{\vx}_{i} = \frac{1}{k+1}\sum_{j = 1}^{k+1} \vec{\widehat{d}}_{i,j},
\end{align}
where we recompute all $\wh{\vx}_i$ as they are used in the $k+1$ iteration.
To see why this works, we separate the sum as
\begin{align*}
\frac{1}{k+1}\sum_{j = 1}^{k+1} \vec{\widehat{d}}_{i,j} =  \vx_{i} - \frac{1}{k+1}\sum_{j = 1}^{k+1} \vx_{j} + \frac{1}{k+1}\sum_{j = 1}^{k+1} \vec{\nu}_{i,j}.
\end{align*}
We observe that $-\frac{1}{k+1}\sum_{j = 1}^{k+1} \vx_{j}$ is the same translation for all points $\wh{\vx}_i$. The consequence of this approach is that the total noise is reduced as we average its different realizations over $k+1$ values. Note that since Algorithm~\ref{algo:support_recovery} assumes that $\wh{\vx}_{1} = \vec{0}$ in $ \wh{\calX}_k$, we also translate back all the points by $-\wh{\vx}_{1}$ after each update.

Unfortunately, the idea of caching the differences introduced in Section~\ref{sec:complexity_analysis} is not compatible with the denoising of the partial solutions. As at each step we modify the partial solution set $\wh{\calX}_k$, the differences between $\wh{\calX}_k$ and $\widetilde{\calD}$ change accordingly, which makes it impossible to cache them. Hence, there exists a hard trade-off between quality and complexity, and we should pick the right strategy depending on the requirements of each specific practical scenario.

\subsection{Comparison of improvement strategies}
\label{sec:comparison_improvements}

Last, we evaluate the significance of our proposed improvements on Algorithm~\ref{algo:support_recovery}. We quantify the results using the index-based error introduced in Section~\ref{sec:performance_analysis}, as well as the $\ell^2$ error, which we define as the $\ell^2$-norm of the difference between the underlying points $\calX$ and their estimation $\wh{\calX}$.\footnote{This requires to first align the two sets of points $\calX$ and $\wh{\calX}$ by minimizing the $\ell^2$-norm between their elements, subject to any shift and/or reflection.}

The comparison of the different improvement strategies is illustrated in Fig.~\ref{fig:denoising_comparison}. In this experiment, we draw $K=6$ one-dimensional points uniformly at random from the interval $[0,1]$ and add Gaussian noise $\mathcal{N}(0, \sigma^2)$ on their pairwise differences. We run Algorithm~\ref{algo:support_recovery} and the proposed improvements for different noise levels $\sigma$.
It is clear that all the proposed strategies enhance the original algorithm, with respect to both the index-based error and the $\ell^2$ error.

Moreover, we also observe that different strategies combine constructively: for example, the symmetric cost function decreases the $\ell^2$ error by $5\%$ on average, while deleting solutions from the set of differences improves the results by $27\%$ on average. When combined together, the average error decreases by $59\%$. Including the denoising further enhances the algorithm, as the average error decreases by $62\%$.
Similarly, for the index-based error there is an evident shift between the phase transitions of the original algorithm with and without improvements.

\section{Influence of the points locations}
\label{sec:star}

The algorithm performance around the phase transition in Fig.~\ref{fig:phase_transition} also seems to indicate that some configurations of points are easier to recover than others. In this section, we run a small experiment to visualize the challenges posed by certain configurations. 

\begin{figure}[tb!]
\centering
    \includegraphics[width=1\linewidth]{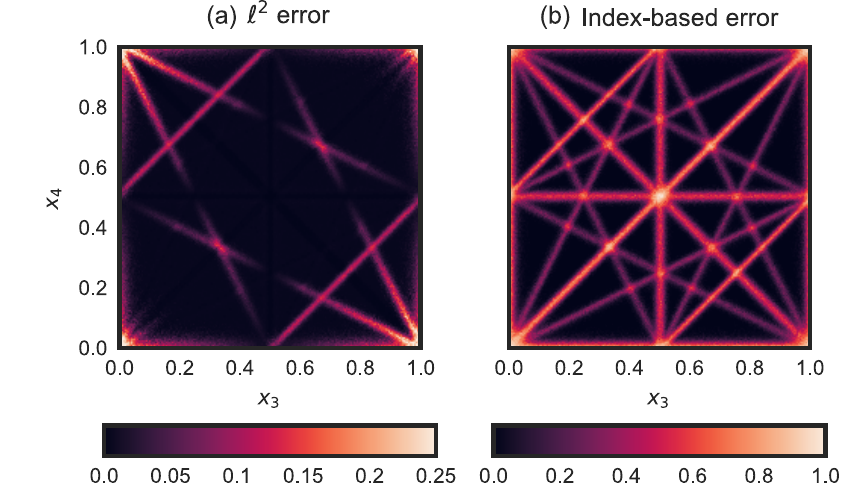}
    \caption{Influence of the points' locations on the estimation errors. We solve a $1$D instance of the problem with $K=4$, $x_1=0$, and $x_2=1$. The locations $x_3$ and $x_4$ vary along the $x$- and $y$-axis.
    \label{fig:star}
    }
\end{figure}

We consider a low-complexity setup ($K = 4$, $D=1$), fix the support boundaries, that is $x_1 = 0$ and $x_2 = 1$, and study the reconstruction error for various pairs $(x_3, x_4) \in [0,1]^2$. We generate several instances of this problem and perturb the differences in $\calD$ with additive Gaussian noise with zero mean and $\sigma = 0.01$. We measure the performance of Algorithm~\ref{algo:support_recovery} 
(with all the improvements introduced in Section~\ref{sec:improvements}) using both the index-based and the $\ell^2$ error. The average errors are then shown in Fig.~\ref{fig:star}, where we observe that there exist some combinations of points that lead to a significantly higher error.

We now develop intuition about a few interesting cases that emerged from the previous experiment. For the sake of simplicity, we consider a noiseless setting where collisions in the ACF or non-uniqueness of the solution are the only causes of challenging configurations.

\begin{enumerate}
    \item \emph{Collision between a difference and a point.} When a difference and a point collide, it can happen that the difference is mistaken for the point. This does not influence the $\ell^2$ error, but causes an index-based error. An example of such a case is when $x_3 = x_4$ (the main diagonal in Fig.~\ref{fig:star}): both the difference $x_4-x_3$ and $x_1$ have value $0$. As a consequence, the sets
    $\calX' = \{x_1, x_2, x_3, x_4 \}$ and  
    $\calX'' = \{x_4-x_3, x_2, x_3, x_4 \}$ are both equal to 
    $\calX = \{0, 1, x_3, x_3 \}$, but the latter is not of the form~\eqref{eq:solution_form}.

    \item \emph{Constant difference 0.5}. When $x_4=x_3 \pm 0.5$, we can actually find more than one set of $4$ points that map to a subset of the given differences. In the case $x_4=x_3 + 0.5$, the differences are $\calD = \pm \{ 0, 1, x_3, x_3+0.5, 1-x_3, 0.5-x_3, 0.5\}$; thus, $\calD$ contains all pairwise difference from both $\calX' = \{0, 1, x_3, x_3 +0.5 \}$ and $\calX'' = \{0, 1, 0.5, x_3 \}$. However, $\calX''$ does not lead to a zero $\ell^2$ error.  
    
    
    \item \emph{Collision of differences when adding a new point to the solution set.} This is for example the case of $x_4 = 1 -2x_3$ with $\calD = \pm \{ 0, 1, x_3, 1 - 2x_3, 1-x_3, 2x_3, 1-3x_3\}$. The differences $0$ and $1$ are always selected in the first and the second step. In the third step, we could potentially add $2x_3$ to $\calX_2 =\{0, 1 \}$ and reduce the set of differences to $\calD = \pm \{ x_3, 1-x_3, 1-3x_3\}$. Next, we select $x_3$ as a new point. We can verify that the differences of $x_3$ and the values in $\calX_3  =\{0, 1, 2x_3 \}$ exist in $\calD$. However, in this verification we use the value $x_3$ in $\calD$ twice: once as the difference between $x_3$ and $0$, and once as the difference between $x_3$ and $2x_3$. The set of pairwise differences of $\calX_4  =\{0, 1, 2x_3, x_3 \}$ is indeed contained in the original $\calD$, but neither its $\ell^2$ error nor its index-based error is zero.  
    Notice that if we swap the third and the fourth step, this confusion would be avoided as $x_3$ would be removed from the set of differences in the third step.
    
\end{enumerate}

These three cases explain all the segments visible in Fig.~\ref{fig:star}. Such an analysis also applies to noisy regimes; the main difference is that we move from very localized configurations to blurrier areas where the solution is ambiguous. In fact, we introduced some noise into the experiment in Fig.~\ref{fig:star} to enable the visualization of the \emph{lines} identifying challenging patterns---a noiseless setting would have just led to infinitesimally thin lines. Such patterns become blurrier and wider as noise increases. 
These areas where reconstruction is harder also explain the not-so-sharp phase transition in Fig.~\ref{fig:phase_transition}: when drawing supports of $K$ elements at random, the probability of hitting a challenging pattern significantly grows with the noise. To the limit, these blurred lines cover the entire domain and the probability of success is null.

\section{Comparison with Charge Flipping}
\label{sec:charge_flipping}

\begin{figure*}[t!]
\centering
    \includegraphics[width=\textwidth]{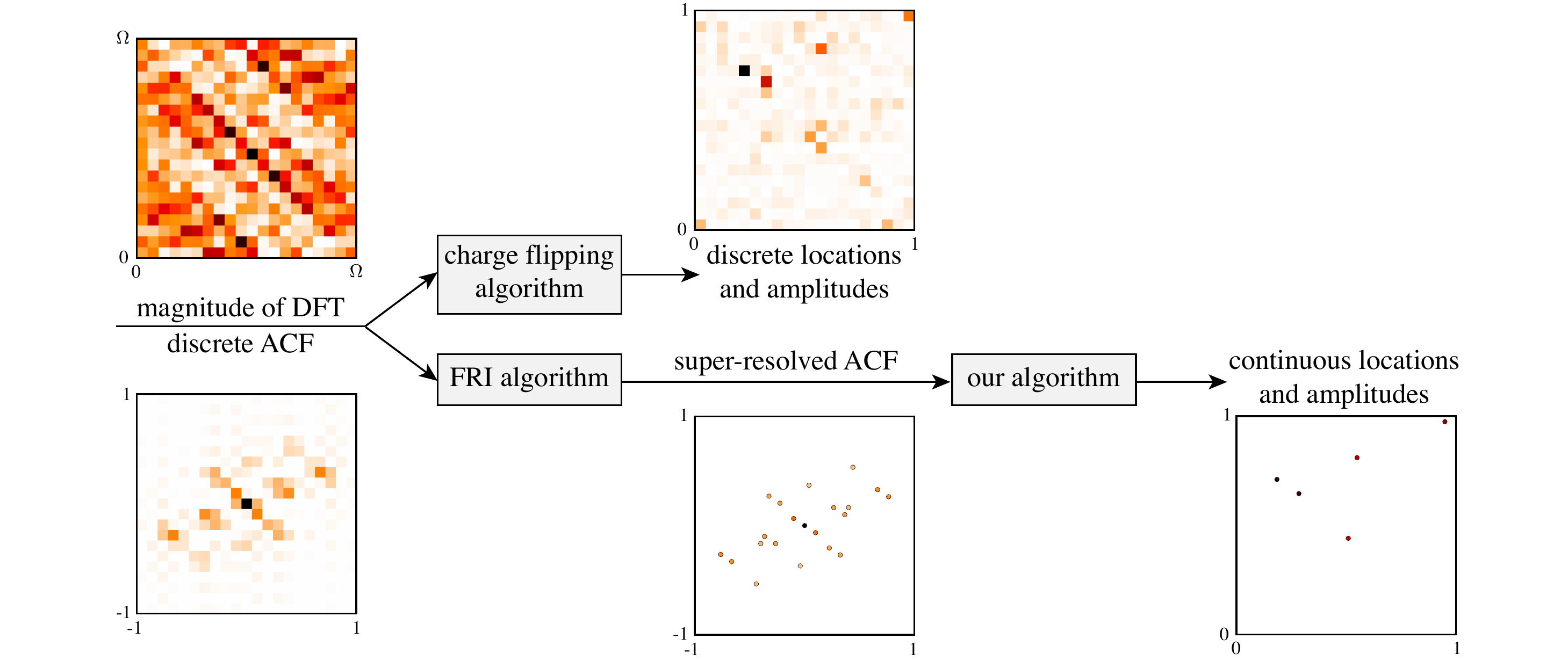}
    \caption{PR pipeline for Charge Flipping and our algorithm. First the signal $a(\vx)$ is sampled and we observe the magnitude of its DFT, $A_{\vec{m}}$, which also corresponds to a discrete version of its ACF. These DFT coefficients are directly used by Charge Flipping to recover a discretized support of $f(\vx)$. Our approach proceeds in two stages: first, using FRI we compute a super-resolved version of the ACF, and then by applying the proposed algorithm, we recover the continuous version of $f(\vx)$.
    \label{fig:schema}
    }
\end{figure*}

In this section, we evaluate the performance of the proposed PR algorithm in comparison with other state-of-the-art methods. Recall that our algorithm is, to the best of our knowledge, the first to operate in a continuous-support setup, whereas other algorithms assume discrete signals.
Indeed, the vast majority of PR methods are simply not designed to work with continuous supports; examples are PhaseLift~\cite{Candes2015phase}, which recasts the PR problem as a semi-definite program, and GESPAR~\cite{Shechtman2013}, which linearizes the PR problem with the damped Gauss-Newton method. In general, these approaches assume that the signals of interest are sparse vectors. As seen in Fig.~\ref{fig:sampling_scheme}, when the locations are not aligned with the sampling grid, the discretized signal contains very few---if any---nonzero entries as the scattering function spreads the sharp continuous locations.

A few algorithms can be adapted to work with continuous supports, but they fall short when the measured support $\wt{\calD}$ is noisy. This is the case of TSPR~\cite{JaganathanOH13}, which relies on the triangle equality between locations to recover the support; as soon as the locations are corrupted with noise, these equalities do not hold anymore. 

The closest point of comparison to our method is the \emph{Charge Flipping} algorithm~\cite{Oszlanyi:2004gb,Oszlanyi:2008fx}; even though it operates in a discrete domain, our experiments have shown that it is resilient to some noise on the ACF support.

\subsection{Charge Flipping}
Charge Flipping is one of the reference algorithms in crystallography. It belongs to the class of dual-space algorithms as it alternatively acts on the spatial and Fourier domains, designated \emph{real} and \emph{reciprocal} space in crystallography. After randomly assigning a phase to the observed magnitudes of the discrete Fourier transform (DFT) coefficients, it iteratively performs the following two operations:
\begin{enumerate}
    \item In the real space, it flips the sign of the values that are below some fixed threshold $\delta$.
    \item In the reciprocal space, it enforces that the magnitude of the signal corresponds to the measured magnitude. 
\end{enumerate}

Charge Flipping directly takes as input the DFT coefficients of the ACF, while our support recovery algorithm operates on a continuous version of the ACF. This is a significant advantage of our algorithm over Charge Flipping as we do not assume that the support of the points is aligned with a grid. To have an adequate comparison between the two, we need to consider the entire pipeline, combining the three algorithms exposed in Section~\ref{sec:algorithms}; this is illustrated in Fig.~\ref{fig:schema}.


\subsection{Experimental setup and results}

We generate DFT coefficients corresponding to a sparse signal as described in~\eqref{eq:sparse_model}, discard their phase information, and corrupt them with zero-mean Gaussian noise. Notice that in Sections~\ref{sec:performance_analysis} and~\ref{sec:improvements}, we assume noise on the support of the points; here, we are dealing with noise that is applied to the DFT coefficients instead. Obviously, these noisy DFT coefficients also lead to a noisy support of the super-resolved ACF, but it is not Gaussian anymore. In fact, as FRI algorithms rely on nonlinear methods, the noise of its output is difficult to characterize.

The discretization of the Fourier domain is equivalent to a periodization of the spatial domain. As a consequence, the squared magnitude of the DFT coefficients corresponds to a \emph{circular} ACF. While it is certainly possible to adapt Algorithm~\ref{algo:support_recovery} to handle circular ACFs by testing all the possible $2^D$ quadrants for every observed $\wt{\vd} \in \wt{\calD}$, we chose to zero-pad the support of $f(\vx)$ until its ACF is not circular anymore.

Regarding Charge Flipping, we notice that its performance highly depends on the initial solution as well as the choice of $\delta$. To avoid giving an unfair advantage to our algorithm, we run Charge Flipping $10$ times for each experiment and pick the best solution; practical experiments show that the performance gain is marginal when going above such a number of repetitions. Furthermore, best practice~\cite{Oszlanyi:2008fx} suggests to pick $\delta = b \theta$, where $b$ is a constant around $1$-$1.2$ and $\theta$ is the standard deviation of the measured signal. Our experiments showed that progressively decreasing the value of $\delta$ also improves the performance of Charge Flipping. This mimics the behavior of the simulated annealing algorithm, where the temperature is steadily decreased until convergence.

\begin{figure}[tb!]
\centering
    \includegraphics[width=1\linewidth]{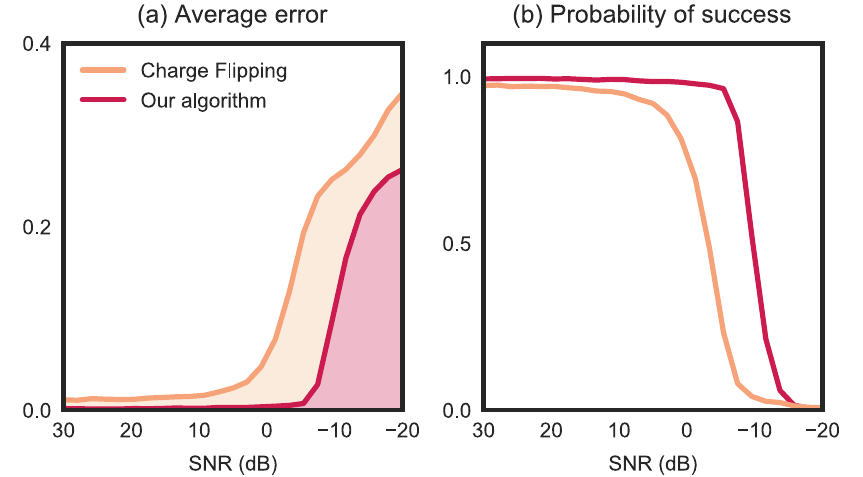}
    \caption{Comparison of our algorithm with Charge Flipping. The performance is evaluated for $K=5$ $1$D points chosen uniformly at random from $[0, 1]$. The number of DFT coefficients is $200$. Figure (a) shows the $\ell^2$ reconstruction error on the locations for different values of the input SNR. Figure (b) reports the percentage of success: we consider that the algorithms fail when the resulting $\ell^2$ error is larger than some threshold $0.04$. 
    \label{fig:charge_flipping}
    }
\end{figure}

Then, given noisy DFT coefficients as input, we compare the $\ell^2$ error on the support of the points for both algorithms, as well as a probability of successfully recovering the support.
To define the latter, we say that an algorithm fails when the $\ell^2$ error is higher than a specific threshold. Fig.~\ref{fig:charge_flipping} shows that our FRI super-resolution algorithm surpasses Charge Flipping in terms of both metrics. 
It is not surprising that our algorithm exhibits a superior performance in a low noise regime---it even achieves exact reconstruction in the absence of noise---since it is not bound to a grid.
On the other hand, Charge Flipping always suffers from approximation errors due to the implicit discretization: in the noiseless case and for a grid of size $200$, we calculate that the expected $\ell^2$ error on the support of $K=5$ points is about $0.0056$, which is in adequacy with the baseline observed in Fig.~\ref{fig:charge_flipping}a.
Simulations also indicate that our algorithm outperforms Charge Flipping in high noise environments. Indeed, the reconstruction error is consistently lower and its phase transition compares favorably as well.

\section{Conclusion}

We presented a novel approach to solve the phase retrieval problem for sparse signals.
While conventional algorithms operate in discretized space and recover the support of the points on a grid, the power of FRI sampling combined with the sparsity assumption on the signal model enables to recover the support of the points in continuous space. We provided a mathematical expression that approximates the probability of success of our support recovery algorithm and confirmed our result via numerical experiments. We observed that while our algorithm runs in polynomial time with respect to the sparsity number of the signal, it remains relatively costly.
To alleviate the computational costs without impacting the quality of the reconstruction, we proposed a caching layer to avoid repeating calculations. Furthermore, we introduced several improvements that contribute to enhance the quality of estimation in the presence of noise. Finally, we showed that our super resolution PR algorithm outperforms Charge Flipping, one of the state-of-the-art algorithms, both in terms of average reconstruction error and success rate.


%



\section*{Acknowledgment}

The authors would like to thank Adam Scholefield for his comments and advice on the manuscript.

\ifCLASSOPTIONcaptionsoff
  \newpage
\fi



\bibliographystyle{IEEEtran}
\bibliography{biblio}

%

\begin{IEEEbiography}{Gilles Baechler}
Biography text here.
\end{IEEEbiography}

\begin{IEEEbiography}{Miranda Krekovi\'c}
Biography text here.
\end{IEEEbiography}

\begin{IEEEbiography}{Juri Ranieri}
Biography text here.
\end{IEEEbiography}

\begin{IEEEbiography}{Amina Chebira}
Biography text here.
\end{IEEEbiography}

\begin{IEEEbiography}{Yue M. Lu}
Biography text here.
\end{IEEEbiography}

\begin{IEEEbiography}{Martin Vetterli}
Biography text here.
\end{IEEEbiography}




\end{document}